\documentclass[a4paper,fleqn]{cas-sc}

\usepackage[authoryear]{natbib}
\def\tsc#1{\csdef{#1}{\textsc{\lowercase{#1}}\xspace}}
\tsc{WGM}
\tsc{QE}
\tsc{EP}
\tsc{PMS}
\tsc{BEC}
\tsc{DE}

\newcommand\arcsec{\mbox{$^{\prime\prime}$}}%

\begin{document}
\let\WriteBookmarks\relax
\def\floatpagepagefraction{1}
\def\textpagefraction{.001}

\shorttitle{Sequential Fragmentation of Comet C/2025 K1 (ATLAS)}

\shortauthors{D. Bodewits et~al.}

\title [mode = title]{Sequential Fragmentation of C/2025 K1 (ATLAS) After Its Near-Sun Passage}                      

%
\author[1]{D. Bodewits}[type=editor,
                        auid=000,
                        orcid=0000-0002-2668-7248]

\cormark[1]

\fnmark[1]

\ead{dennis@auburn.edu}


\credit{Conceptualization of this study, Methodology, Manuscript Development}

\affiliation[1]{organization={Auburn University},
    addressline={Physics Department, Edmund C. Leach Science Center}, 
    city={Auburn},
    postcode={36849}, 
    state={AL},
    country={USA}}

\author[1]{J. W. Noonan}[type=editor,
                        auid=000,
                        orcid=0000-0003-2152-6987]
\credit{Data Reduction and Analysis, Methodology, Software, Manuscript Development}
\ead{noonan@auburn.edu}

\author[2]{M. S. P. Kelley}
   [orcid=0000-0002-6702-7676]
\credit{Data Reduction and Analysis}
\ead{msk@astro.umd.edu}

\affiliation[2]{organization={University of Maryland},
    addressline={Department of Astronomy}, 
    city={College Park},
    postcode={20742}, 
    state={MD},
    country={USA}}

\author[3]{C. E. Holt}
   [type=editor, orcid=0000-0002-4043-6445]
\fnmark[2]
\fntext[2]{LSST-DA Catalyst Fellow}
\ead{cholt@lco.global}
\credit{Observations}

\affiliation[3]{organization={Las Cumbres Observatory},
    addressline={6740 Cortona Drive Suite 102}, 
    city={Goleta},
    postcode={93110}, 
    state={CA},
    country={USA}}

\author[3]{T. A. Lister}
    [type=editor, orcid=0000-0002-3818-7769]
\ead{tlister@lco.global}
\credit{Observations}

\author[4]{H. Usher}
   [type=editor, orcid=0000-0002-8658-5534]
\ead{helen.usher@open.ac.uk}
\credit{Observations}

\affiliation[4]{organization={School of Physical Sciences, Open University},
    city={Milton Keynes},
    postcode={MK7 6AA}, 
    country={UK}}
 \affiliation[4]{organisation={School of Physics and Astronomy, Cardiff University},
    addressline={Queens Buildings, The Parade},
    city={Cardiff}, 
    postcode={ CF24 3AA}, 
    country ={UK}}

\author[5]{C. Snodgrass}
   [type=editor, orcid=0000-0001-9328-2905]
\ead{csn@roe.ac.uk}
\credit{Manuscript Development}
\affiliation[5]{organization={University of Edinburgh},
    addressline={Royal Observatory}, 
    city={Edinburgh},
    postcode={EH9 3HJ}, 
    country={UK}}

\author[6]{B. J. R. Davidsson}
   [type=editor, orcid=0000-0002-8725-6644]
   \credit{Manuscript Development}
\ead{Bjorn.Davidsson@jpl.nasa.gov}

\affiliation[6]{organization={Jet Propulsion Laboratory, California Institute of Technology},
    addressline={M/S 183-601, 4800 Oak Grove Drive}, 
    city={Pasadena},
    postcode={91109}, 
    state={CA},
    country={USA}}

\author[7,8]{S. Greenstreet}
   [type=editor, orcid=0000-0002-4439-1539]
\credit{Instrument development}
\ead{sarahjg@uw.edu}

\affiliation[7]{organization={NSF-DOE Vera C. Rubin Observatory/NSF NOIRLab},
    addressline={950 N. Cherry Ave.}, 
    city={Tucson},
    postcode={85719}, 
    state={AZ},
    country={USA}}

\affiliation[8]{organization={DiRAC Institute and the Department of Astronomy, University of Washington},
    addressline={3910 15th Ave. N}, 
    city={Seattle},
    postcode={998195}, 
    state={WA},
    country={USA}}

\cortext[cor1]{Corresponding author}

\begin{abstract}
Comet C/2025 K1 (ATLAS) reached perihelion at 0.33~au on 2025 October 8. Daily monitoring by the LCO Outbursting Objects Key Project revealed a major activity increase between November 2 and 4, accompanied by rapid changes in coma morphology. Serendipitous HST/STIS acquisition images obtained on November 8–10 captured the comet only days after this event and resolved five fragments, providing an early high-resolution view of a nucleus in the process of disruption. Fragment motions and morphologies indicate a hierarchical fragmentation sequence, including a slow secondary split of fragment II. Back-extrapolation shows that both the primary and secondary breakups preceded their associated photometric outbursts by roughly one to three days. This consistent lag, together with the appearance of thin, short-lived arclets around fragment I in the first HST epoch, suggests that freshly exposed interior material warms rapidly but requires time before dust can be released efficiently. { Given the comet’s close perihelion passage, rotational instability driven by enhanced outgassing torques is a plausible contributor to nucleus disintegration and dust release, and may represent the primary source of the observed brightening.} These combined ground- and space-based observations provide rare, time-resolved constraints on the thermal and structural evolution of a fragmented comet near perihelion and highlight the scientific value of capturing a nucleus within days of disruption { when thermal adjustment, dust mantle re-formation, and outgassing-driven torques jointly govern the onset of activity.}
\end{abstract}


\begin{highlights}
\item HST resolved a hierarchical fragmentation sequence in C/2025~K1 (ATLAS) within days of disruption, while the LOOK survey provided continuous photometric context and tracked the subsequent evolution of its activity.
\item Fragment kinematics and a consistent 1–3~day delay between breakup and brightening show that post-fragmentation activation is controlled by { thermal adjustment of newly exposed surfaces}, analogous to delayed activity following \textit{Rosetta}’s cliff collapses on 67P.
\item Short-lived, shell-like arclets observed around fragment~I in the first HST epoch indicate dust-shell release during initial warm-up rather than gas-driven pressure waves, revealing the physical processes governing early post-fragmentation morphology and the onset of efficient dust production.

\end{highlights}


\maketitle

\section{Introduction}

Comet C/2025~K1 (ATLAS) reached perihelion at 0.33~au on 2025~October~8 and subsequently
experienced a sequence of outbursts culminating on November~4. Pre–perihelion spectra
revealed an extreme depletion in carbon–bearing species, yielding one of the lowest CN/OH
ratios ever measured, comparable only to a few chemically anomalous comets that have
been interpreted as possible extrasolar candidates \citep{Schleicher2008,Schleicher25,Manzini2025a}.
Fragmentation is a common evolutionary end state of long-period comets
\citep{Boehnhardt2004}, frequently associated with progressive mass loss, thermal stresses, and
rotational spin-up driven by asymmetric outgassing \citep{Jewitt2004,Jewitt2022}. Because the
interior of a comet is rarely exposed, fragmentation events provide valuable opportunities to
probe subsurface structure, volatile reservoirs, and short-timescale activation processes.

Following perihelion, C/2025~K1 was monitored closely by the Las Cumbres Observatory Outbursting Objects Key
(LOOK) Project \citep{lister22}. Comparison of the observations on November~4 with
Zwicky Transient Facility (ZTF) photometry showed that the comet underwent at least a 0.9~mag outburst between
November~2 and~4, accompanied by the rapid development of a two-component coma: an extended
parabolic outer feature and a compact oblate inner coma of
freshly released material \citep{Kelley2025}. Independent small-telescope photometry (A.~Pearce,
priv.~comm.) places the onset between November~3~10{:}33~UT and November~4~10{:}33~UT,
and over the following nights the central condensation evolved to show multiple brightness
peaks as fragments drifted far enough apart to be spatially separated from the ground \citep{MPEC_K25W40_2025, MPEC_K25W56_2025}.

Serendipitous {Hubble Space Telescope} (HST) observations, obtained only days after
this outburst, captured the nucleus in the act of breaking apart and resolved at least five
fragments \citep{Noonan2025}. Their relative motions and evolving morphologies reveal a hierarchical fragmentation sequence and provide one of the earliest high-resolution records of a near-Sun cometary disruption. As their projected separations increased, some of these fragments became detectable in independent ground-based imaging.

In this manuscript we combine HST imaging with contemporaneous LOOK Project photometry to characterize the morphology, motion, and activity of the fragments, place the breakup within the comet’s broader brightness evolution, and assess whether fragmentation exposed compositionally distinct interior material.

\section{Observations}
\begin{figure}
	\centering
		\includegraphics[width=0.75\textwidth]{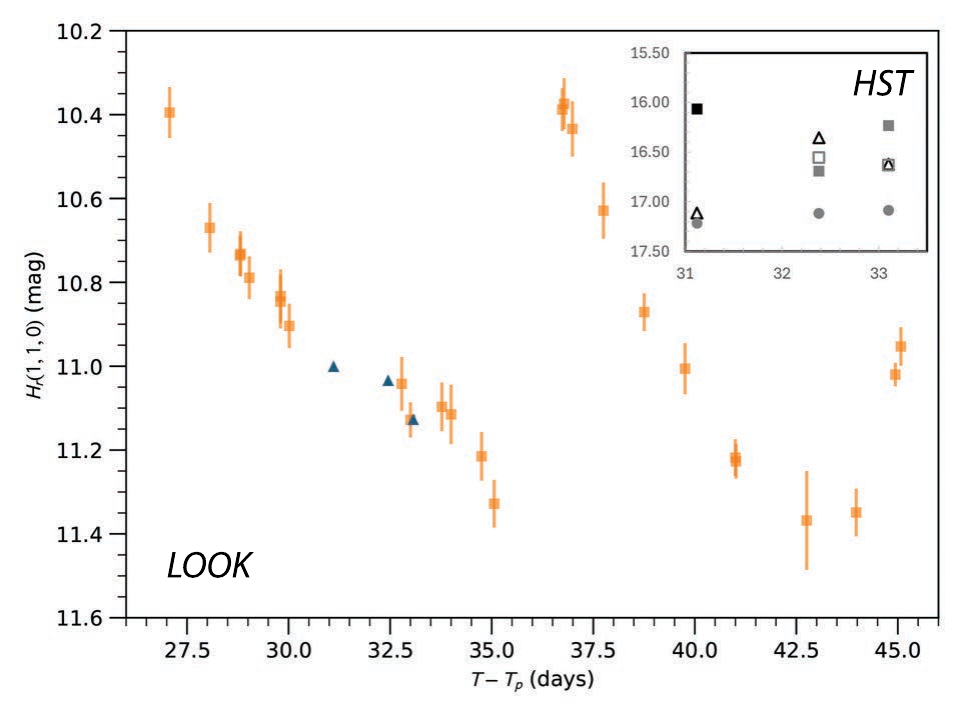}
	\caption{
LOOK r-band photometry of C/2025~K1 (ATLAS) measured in a circular aperture of 10{,}000~km in radius (orange squares).HST/STIS photometry within a similarly-sized aperture were bootstrapped to the Nov.~10 LOOK data and are depicted by blue triangles.  The inset panel presents the absolute magnitudes of the individual fragments {a within apertures with radii of 0.15~arcsec} obtained from HST/STIS (cf.\ Fig.~\ref{FIG:HST}): open triangles denote fragment~I; black, open, and grey squares represent fragments II, IIa, and IIb, respectively; grey circles indicate fragment~IV (magnitudes of IV are decreased by $-2$~mag to aid visualization) { The time on the x-axis is the time since perihelion, which occurred on Oct 8.4, 2025.}
}
	\label{FIG:look_phot}
\end{figure}

\subsection{LCO Outbursting Objects Key Project}
\noindent C/2025 K1 has been monitored by the LCO Outbursting Objects Key (LOOK) Project \citep{lister22} since 2025 June 3 with a break from 2025 Sep 2 to 2025 Nov 3 during perihelion passage and low solar elongation. Since observations resumed on 2025 Nov 4, C/2025 K1 was monitored near daily using the LCO 1-m telescopes and Sinistro instruments (0.39\arcsec{} pixel scale, 4k$\times$4k CCDs) in SDSS g, r and i filters. Observation requests switched to visits every $\sim8$ hours from 2025 Nov 10. Observations from a single visit were processed to remove instrumental artifacts (bias and flat-fielding), produce source catalogs, and provide an astrometric calibration with the BANZAI pipeline \citep{McCully2018}.  Data were downloaded from the LCO Science Archive and automatically processed by custom LOOK Project software to photometrically calibrate the data to the PS1 system \citep{lister22}, and combine images into per-filter stacks for analysis.  We measured the brightness of the fragmented comet in 10,000~km radius apertures centered on fragment A.

\begin{figure}
	\centering
		\includegraphics[width=0.3\textwidth]{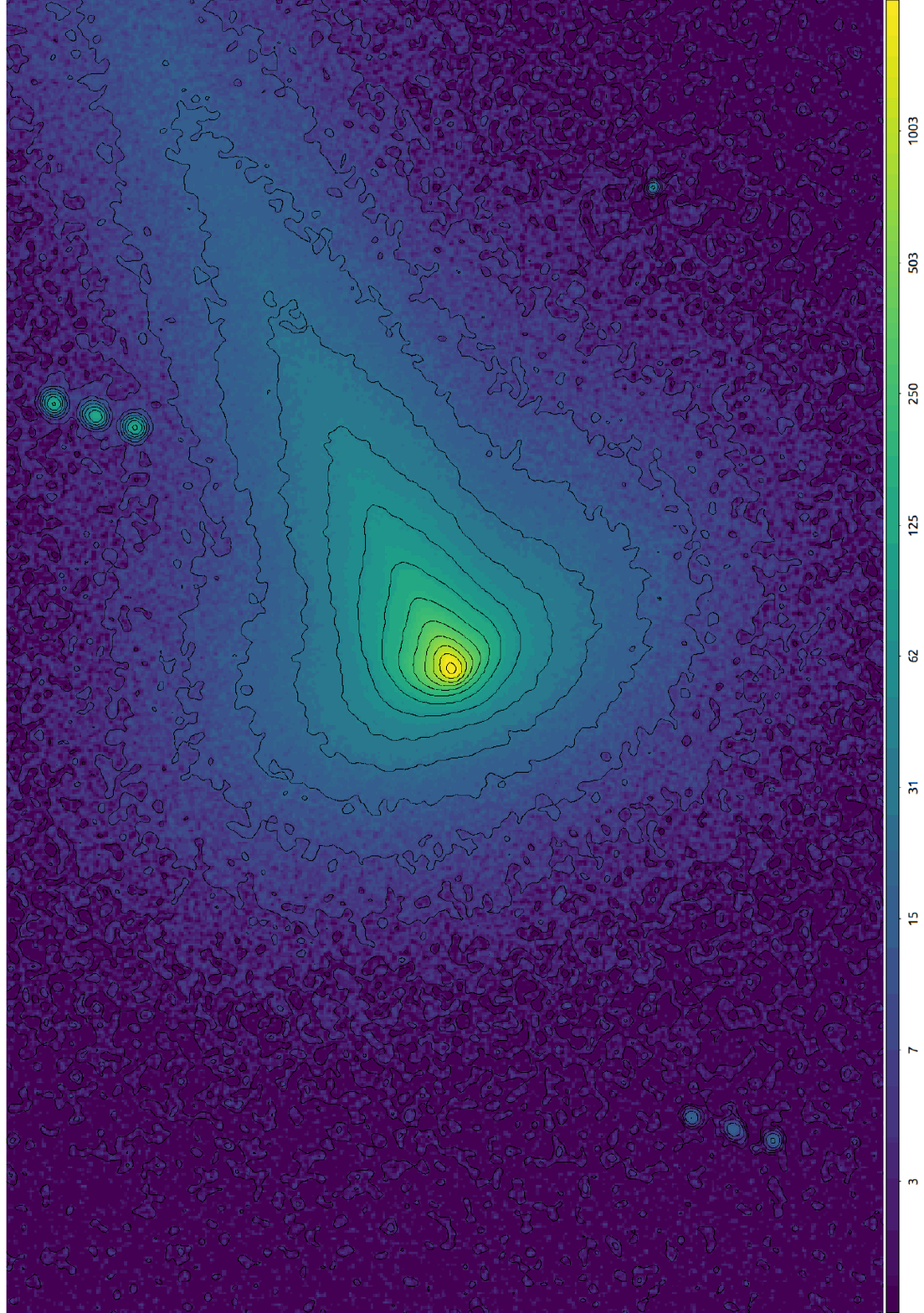}
        \includegraphics[width=0.3\textwidth]{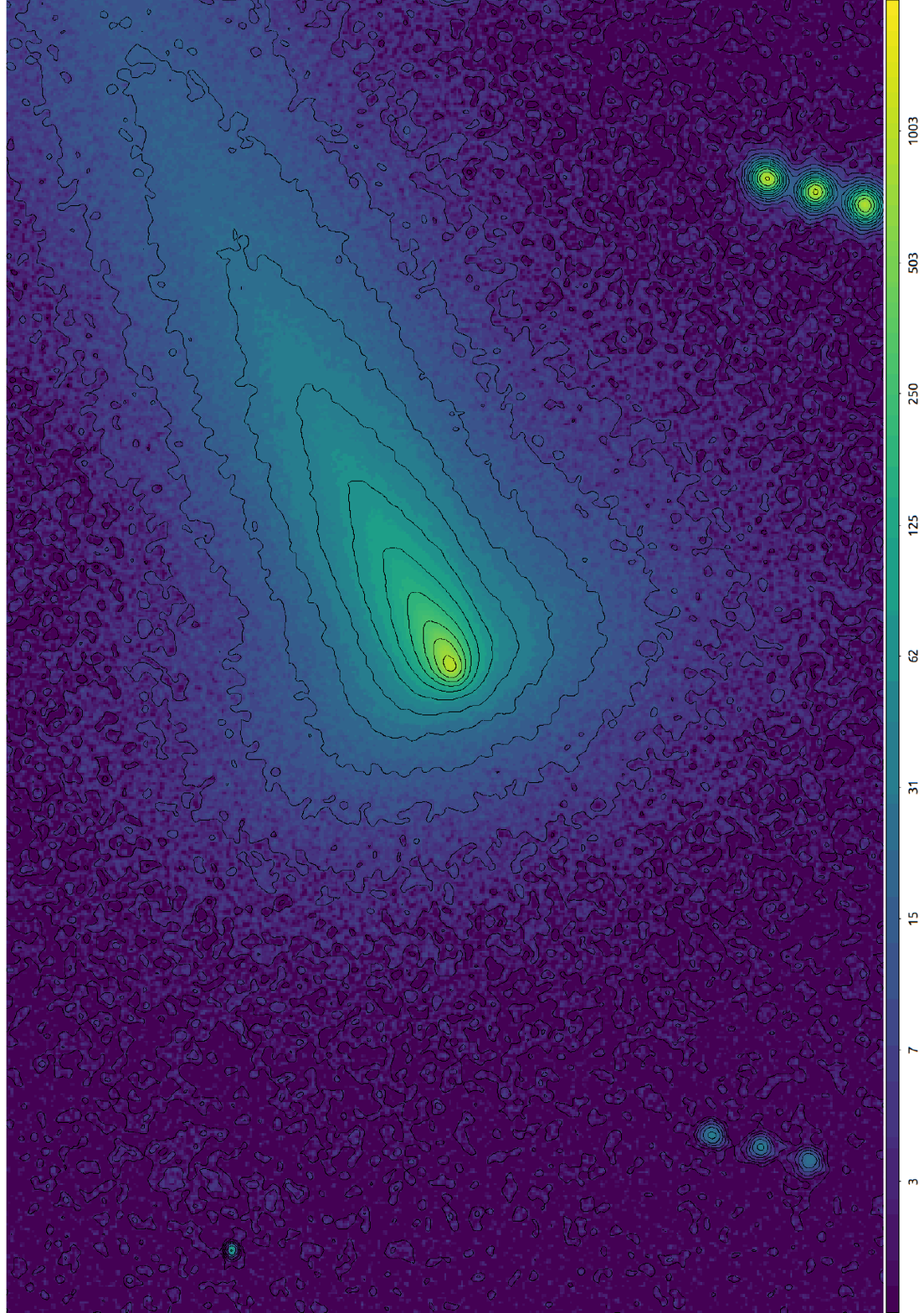}
        \includegraphics[width=0.3\textwidth]{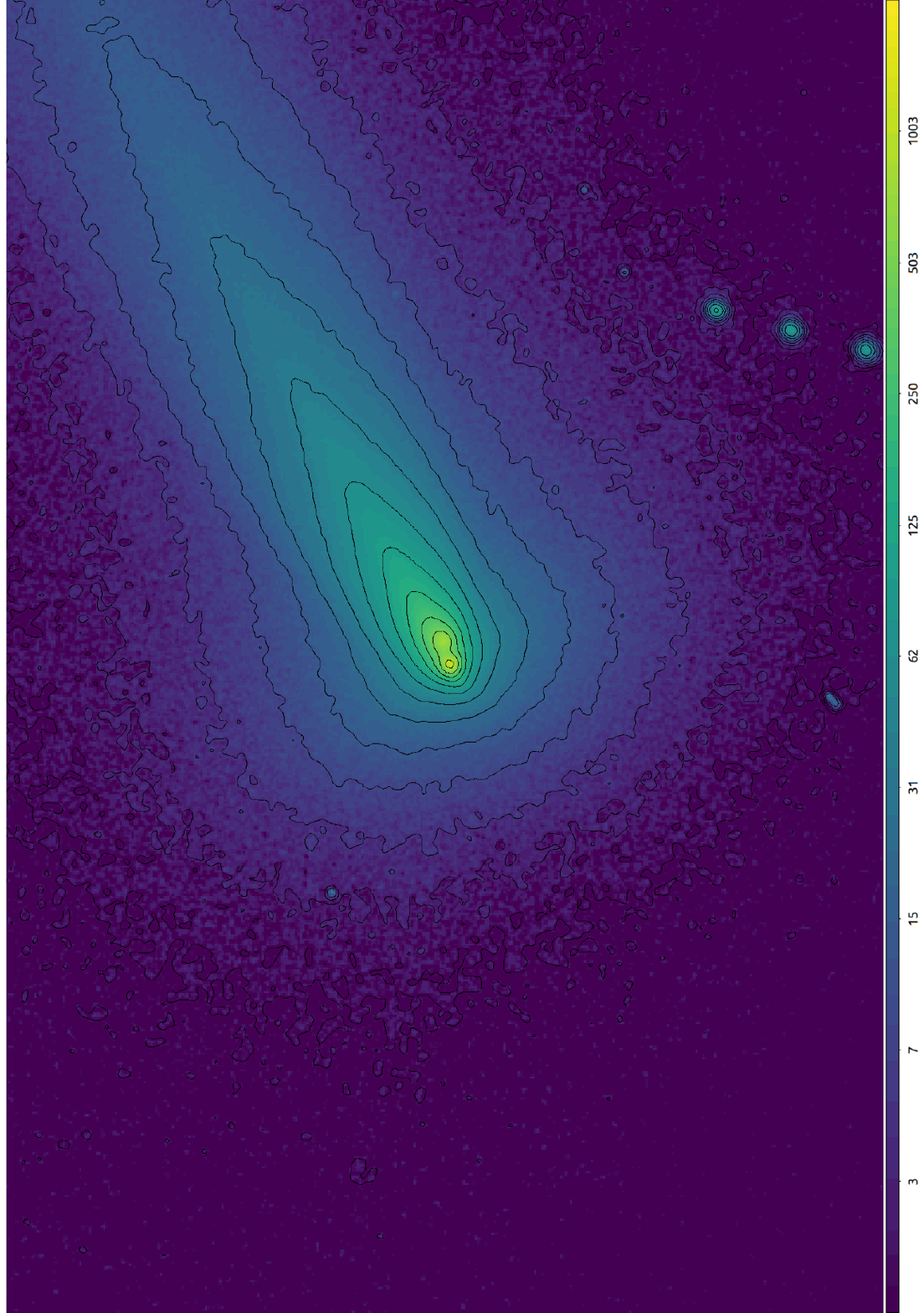}
	\caption{LCO images r-band acquired on Nov. 4.51 (left; after large outburst), Nov. 7.46 (center; 1.1 day before the first HST observation), and Nov. 10.23 (right; 7.5 hr before the last HST observation). The stretch is logarithmic and the color scale is the same in all three images. }
	\label{FIG:LOOK}
\end{figure}

\subsection{HST/STIS}
\begin{table}
    \centering
    \begin{tabular}{clccccc}\toprule
         Number&  Time (UTC)&   T-Tp (days)&r$_h$ (au)&  $\Delta$ (au) &Phase (degrees)& t$_{exp}$ (s)\\\midrule
         1&  Nov. 8, 2025, 13:22&   31.12&0.875&  0.642 &79.9& 20.5\\
         2&  Nov. 9, 2025, 19:42&   32.38&0.901&  0.612 &79.2& 20.5\\
         3&  Nov. 10, 2025, 12:55&   33.10&0.916&  0.597 &78.7& 20.5\\
         \bottomrule
    \end{tabular}
    \caption{Observing log of the HST/STIS MIRVIS images of C/2025 K1 (ATLAS)}
    \label{tab:log}
\end{table}
\noindent The images used here were context frames acquired for a spectroscopic survey using { the Space Telescope Imaging Spectrograph (STIS)} and { the Cosmic Origins Spectrograph (COS)} on board HST (Program~18135, PI~D.~Bodewits). Each consists of a 20.5~s  exposure obtained with the MIRVIS detector and the F28X50LP long-pass filter, whose throughput spans 5500--10{,}000~\AA\ with an FWHM of 2693~\AA\footnote{\url{https://hst-docs.stsci.edu/stisihb/chapter-14-imaging-reference-material/14-3-ccd/ccd-long-pass-imaging-f28x50lp}}. { The radius of the point-spread function  that encloses 80\% of the flux ($R_{80})$$ is \approx$ 0.15~arcsec}. Three pairs of MIRVIS images were obtained over a two-day interval shortly after the comet’s major outburst, providing three well-spaced epochs for tracking the motion, morphology, and evolving activity of the fragments. The observations span 31–33 days past perihelion and cover modest changes in heliocentric and geocentric distance as well as phase angle. A complete observing log, including times, geometry, and exposure details for all three epochs, is given in Table~\ref{tab:log}. These three epochs provide the basis for tracking the motion, morphology, and evolving activity of the fragments.

We measured the brightness of each fragment using a circular aperture with the STIS $R_{80}$ radius of 0.15\arcsec. Photon count rates were converted to fluxes and magnitudes using the calibration constants in the FITS headers \citep{Proffitt2006}, without background subtraction. At the $\sim$0.6~au geocentric distance of the observations, the 0.15\arcsec{} PSF corresponds to a physical scale of $\sim$68~km—far larger than any plausible nucleus—so the extracted fluxes are dust dominated. We therefore applied the Schleicher–Marcus composite dust phase correction\footnote{\url{https://asteroid.lowell.edu/comet/dustphase/table}} to account for the large scattering angles. We also measured the total flux in each frame using a fixed 10{,}000~km-radius aperture centered on the image, again without background subtraction and with the same phase correction. A solar spectrum through the F28X50LP filter is comparable to an $r$-band magnitude, so no color correction was applied. To compare the STIS results with the nearly contemporaneous LOOK photometry on Nov.~10 (within 2.5~hr), we converted both to { absolute magnitudes} $H(1,0,0)$. The LOOK measurement is 0.5~mag brighter than the STIS result, which we attribute to differences in filter bandpasses. We therefore bootstrap the STIS photometry to the LOOK scale by subtracting 0.5~mag from all STIS magnitudes.

\section{Results}
\begin{figure}
    \centering
    \includegraphics[width=\textwidth]{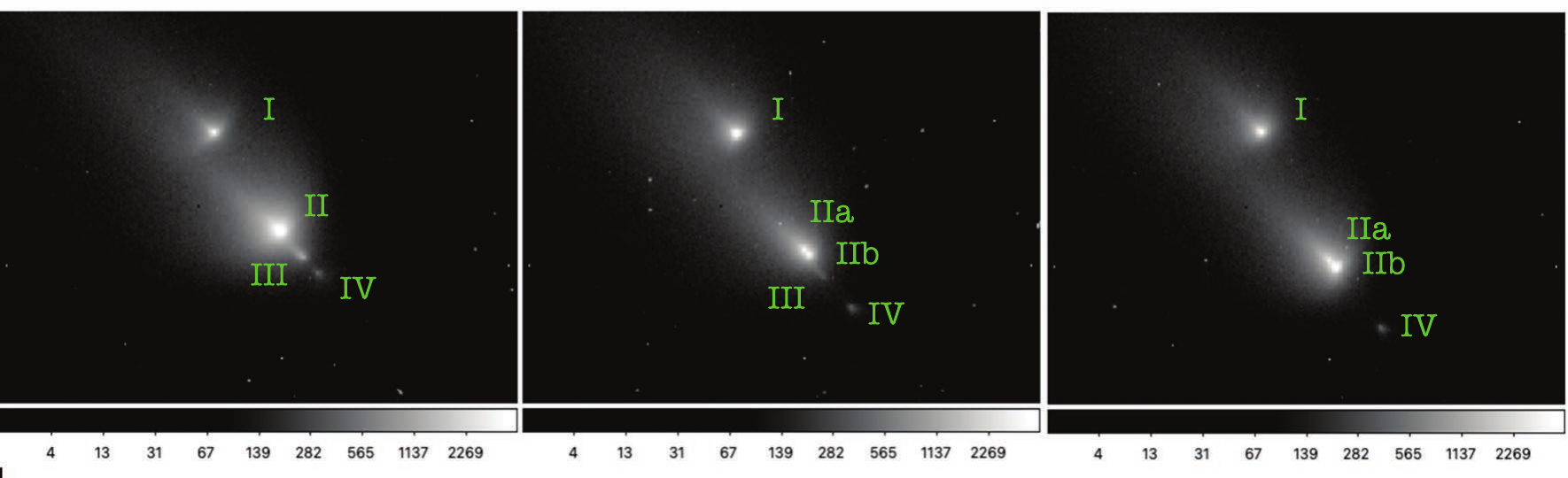}
\caption{
HST/STIS MIRVIS images of C/2025 K1 (ATLAS) obtained on 2025 Nov.~8.56{\ (T$-$T$_p$ = 32.16~days)} (left), Nov.~9.82{ (T$-$T$_p$ = 33.42~days)} (center), and Nov.~10.54{\ (T$-$T$_p$ = 34.14~days)} (right). Each panel uses a logarithmic stretch with the same physical and intensity scale. The field of view is $20\arcsec \times 15\arcsec$, corresponding to roughly $9{,}000 \times 6{,}500$~km at the comet. Fragments are labeled following the notation introduced in the text: the two brightest components (I and II) are present in the first epoch, with fainter fragments III and IV located progressively farther along the projected orbit. By the second epoch, fragment II has separated into components IIa and IIb, and the relative motions and evolving morphologies of all fragments are evident, including arclets around I, a parabolic inner coma around II, the forward fan of IV, and the fading of III. In the final panel, the brightening and coma development of fragment IIb are clearly visible. Grayscale values represent counts in a 21.5~s exposure.
}

    \label{FIG:HST}
\end{figure}
\subsection{Ground-Based Photometry and Context}

\noindent LOOK photometry for Nov.\ 4–22{ [T$-$T$_p$ = 27.61–45.98~days]}, spanning the HST observing window, is shown in Fig.~\ref{FIG:look_phot}. The –0.9~mag outburst of \citep{Kelley2025} is apparent as a decrease in the {magnitude} with time starting with our first observation on November~4~12{:}16~UTC{[T$-$T$_p$ = 27.61~days]}. Two additional outbursts are found in the data, on November~14~04{:}17~UTC{ [T$-$T$_p$ = 36.78~days]} and November~22~09{:}04~UTC{ [T$-$T$_p$ = 45.98~days]}, with strengths of $-0.94\pm0.08$ and $-0.33\pm0.06$~mag for our 10{,}000~km aperture. Figure~\ref{FIG:LOOK} shows selected images of the comet around the time of the HST observations. After the outburst, the comet faded from $H(1,1,0)=10.9$ to 11.0 by November~10{ [T$-$T$_p$ = 33.56~days]}, with small deviations attributable to fragment brightening resolved by HST.

STIS photometry extracted within a 10{,}000~km aperture yield nearly
constant absolute magnitudes of $H=11.0$–$11.1$ across the three
epochs (Nov.\ 8.56–10.54{ [T$-$T$_p$ = 32.16–34.14~days]}; Fig.~\ref{FIG:look_phot}). Although the STIS images show that fragment~IIb was already brightening by Nov.\ 10.53{\ [T$-$T$_p$ = 34.13~days]}, this increase was not immediately apparent in the LOOK photometry because IIa and IIb remained unresolved from the ground. Once their separation increased and IIb’s intrinsic activity strengthened, the effect became visible: between Nov.\ 12~12{:}01 [T$-$T$_p$ = 35.06~days] and Nov.\ 14~05{:}23 [T$-$T$_p$ = 36.78~days], the LOOK data suggest that fragment~IV underwent an $\sim$1.0~mag outburst, producing a newly pronounced parabolic coma clearly distinguishable from the pre-existing coma. This represents the integrated ground-based signature of activity that began at least two days earlier, as captured directly by HST.

A later episode of activity is visible in the LOOK monitoring near the end of the observing window. A new outburst of the eastern fragment occurred between Nov.\ 21~09{:}57{ [T$-$T$_p$ = 45.03~days]} and Nov.\ 22~09{:}04~UTC{ [T$-$T$_p$ = 45.98~days]}, as seen in consecutive r-band frames (logarithmic stretch) obtained on Nov.\ 20, 21, and 22{ [T$-$T$_p$ = 44.03–46.03~days]}. The Nov.\ 22 image shows a newly developed, compact parabolic coma coincident with the eastern fragment, distinct from the broader, fading dust structures associated with the earlier activity. By this date, fragment~I had become increasingly diffuse, consistent with its eventual dispersal into a fading cloud rather than a stable condensation.

\subsection{High-Resolution Morphology and Evolution of the Fragments}

\noindent HST observations directly resolve the evolving activity into individual fragments (Fig.~\ref{FIG:HST}). The first STIS image, acquired on 2025~Nov.~8.56{\ [T$-$T$_p$ = 32.16~days]}, reveals at least four fragments: two brighter components (I and II) separated by 3.4\arcsec, and two fainter components (III and IV) located 4.4\arcsec\ and 5.1\arcsec\ ahead of I along the projected orbit. Fragment~IV shows a distinct forward fan with an opening angle of roughly $100^\circ$, while fragment~I exhibits arclets nearly perpendicular to the dust-tail direction. Fragment~II appears to be the most active, surrounded by a compact parabolic coma.

In the second epoch (Nov.~9.82{ [T$-$T$_p$ = 33.42~days]}; 30.33~hr later), the I--II separation increased to 3.9\arcsec, and II split into two components, IIa and IIb, separated by 0.24\arcsec. Fragment~III evolved from a compact clump into a diffuse smear, while IV remained clearly detectable and had moved to 6.2\arcsec\ ahead of I.
By the final HST epoch (Nov.~10.53{ [T$-$T$_p$ = 34.13~days]}; 17.22~hr later), the separation IIa--IIb increased to 0.30\arcsec, the separation I--IIa had increased to 4.2\arcsec, III had faded below detection, and IV appeared slightly diminished but still present at 6.7\arcsec. Notably, IIb had brightened substantially, indicating the onset of significantly enhanced activity late in the sequence.

The arclets surrounding fragment~I provide an additional diagnostic of its recent activity. Arclet-like features have been reported in other comets \citep[e.g.,][]{Harris1997,Boehnhardt2004}, but in C/2025~K1 they are seen only in the first HST epoch (Nov.~8.56{ [T$-$T$_p$ = 32.16~days]}). Within that epoch they remain tightly centered on I, show no measurable outward expansion over the paired frames, and appear as thin, shell-like structures nearly perpendicular to the dust-tail direction. 

\subsection{Fragment Sizes, Separation Velocities, and the Fragmentation Sequence}

\noindent Using 0.15\arcsec\ (68~km) apertures, the scattering cross sections of the fragments show clear and systematic evolution across the three HST epochs. On Nov.~8.56{\ [T$-$T$_p$ = 32.16~days]}, fragments~I and~II dominate with cross sections of 3.6~km$^{2}$ and 9.4~km$^{2}$, respectively, while fragment~IV is much smaller, contributing only 0.5~km$^{2}$. By the second epoch (Nov.~9.82{\ [T$-$T$_p$ = 33.42~days]}), fragment~II has divided into two pieces: IIa and IIb exhibit comparable cross sections of 6.0~km$^{2}$ and 5.3~km$^{2}$. Fragment~I temporarily brightens to 7.2~km$^{2}$, consistent with the development of arclets, while fragment~IV remains faint at 0.6~km$^{2}$. By the final epoch (Nov.~10.54{\ [T$-$T$_p$ = 34.14~days]}), I settles to 5.6~km$^{2}$, IIa remains steady at 5.6~km$^{2}$, and IIb brightens markedly to 8.1~km$^{2}$, mirroring the onset of enhanced activity visible in the morphology. Fragment~IV is nearly unchanged at 0.6~km$^{2}$. Interpreting these cross sections as upper limits to the effective radii {using the assumptions listed in \citet{Graykowski2019fn}} yields roughly 1.1~km for I, $\sim$1.7~km for II, $\sim$1.3~km for IIa and IIb, and $\sim$0.4~km for fragment~IV.

Projected separation rates further support a hierarchical fragmentation sequence. The I--II separation grows from 3.4\arcsec\ to 4.2\arcsec\ over the two days (Nov.~8.56–10.53{\ [T$-$T$_p$ = 32.16–34.13~days]}), corresponding to $(4.6$--$5.4)\times10^{-6}$~arcsec~s$^{-1}$, or physical speeds of $\sim$3--4~m~s$^{-1}$. Fragments~III and~IV lie progressively farther ahead along the projected orbit, with IV exhibiting the highest apparent motion, consistent with its small size and susceptibility to acceleration. III disperses rapidly, becoming undetectable by Nov.~10{\ [T$-$T$_p$ = 33.56~days]}. The separation of IIa and IIb proceeds much more slowly, at $\sim8\times10^{-7}$~arcsec~s$^{-1}$, indicative of a gentle, near–escape-velocity split.

Extrapolating the projected I--II separation backwards places the primary fragmentation event near 2025~Nov.~1{\ [T$-$T$_p$ = 23.56~days]}, consistent with the timing of the 0.9~mag outburst observed between Nov.~2 and~4{\ [T$-$T$_p$ = 24.56–27.61~days]}. Fragments~III and~IV must have separated earlier and at higher speeds, consistent with their smaller sizes and rapid fading. The secondary breakup of II into IIa and IIb is tightly constrained to the interval between Nov.~8.56 and~9.82{\ [T$-$T$_p$ = 32.16–33.42~days]}, preceding the sharp brightening of IIb that becomes prominent in the final HST frame and in the LOOK survey several days later.

The timing of the morphological changes and the lightcurve implies a 1–3~day delay between fragment exposure and peak dust production. The shell-like arclets around fragment~I reinforce this scenario, consistent with a thermal timescale of several rotation cycles.

We briefly explored an alternative description in which the separations are fit by constant accelerations rather than constant projected speeds. This produces a self-consistent single fragmentation epoch around 2025~Oct.\ 20{\ [T$-$T$_p$ = 11.60~days]}, with higher accelerations for the faint fragments III and IV and lower accelerations for II. This date is consistent, within uncertainties, with an earlier outburst reported by Qicheng Zhang\footnote{\url{https://cometary.org/@qicheng/statuses/01K8BMPX63NCB3K0NXZMYF2XQW}} before post-perihelion LOOK monitoring resumed. However, the limited number of HST epochs, the reliance on projected distances rather than true 3D positions, and the observed secondary breakup of II argue that hierarchical fragmentation provides a more natural and physically plausible description. For these reasons, we adopt the simpler interpretation based on projected separation rates.

\section{Discussion}

\noindent A comparison of the fragmentation chronology with the photometric evolution reveals a systematic delay between the physical breakup of the nucleus and the subsequent rise in brightness. Back–extrapolation of the I--II separation places the primary fragmentation event around 2025~Nov.~1 (within a few days), whereas the major outburst detected in the LOOK and ZTF photometry occurred between Nov.~2.54 and Nov.~4.51~UTC. The delay of roughly $1$--$3$~days suggests that freshly exposed interior ices remained extremely cold immediately after fragmentation and required time for a thermal wave to propagate inward before vigorous sublimation and dust release could commence. The secondary fragmentation of II shows a similar pattern: the IIa--IIb split occurred between Nov.~8.56 and Nov.~9.82, yet IIb brightened substantially only by the final HST epoch on Nov.~10.53, again implying a $\sim$1~day delay. The smaller ($\sim$0.2~mag) outburst detected shortly before the Nov.~10 LCO observations likewise fits within this sequence of delayed activation. These lags are naturally explained through basic thermal physics. For porous cometary material with thermal diffusivity $\kappa \sim 10^{-7}$--$10^{-8}~\mathrm{m^2\,s^{-1}}$ \citep{Davidsson2014}, the penetration depth of the thermal wave is $\delta \sim \sqrt{\kappa t}$. Reaching temperatures of $\sim$180--200~K, sufficient for rapid $\mathrm{H_2O}$ sublimation at $r_h\approx0.9$~au, requires warming layers several centimeters deep. This corresponds to heating timescales of $t \sim 1$--$3$~days, in excellent agreement with the observed photometric delays. Thus, the newly exposed surfaces created by fragmentation require several rotation cycles before effective sublimation can drive strong dust production.

{An alternative, but physically consistent, interpretation of the observed delay is that it reflects the time required to re-establish a mechanically unstable dust mantle on freshly exposed fragment surfaces, rather than solely the time needed to heat subsurface ice to sublimation temperatures. Immediately after fragmentation, the fragment surfaces are expected to expose an intimate mixture of ice and dust representative of the bulk nucleus. Upon solar illumination, such surfaces can begin vigorous gas sublimation on short timescales (hours), but without substantial dust release. Significant dust production may only commence once a porous dust mantle of sufficient thickness has re-formed and thermally decoupled from the underlying sublimation front.}

{ Order-of-magnitude energy-balance considerations support this scenario. For heliocentric distances near $r_h \approx 0.9$~au, the net absorbed solar flux is sufficient to heat newly exposed ice–dust mixtures from $\sim$100~K to water-sublimation temperatures ($\sim$200~K) in only a few hours. However, most of the absorbed energy over the following days is expended on sublimating near-surface water ice, allowing a dust mantle to grow to a thickness of a few centimeters over $\sim$1–3~days. Thermophysical models of comet~67P show that mantles of comparable thickness naturally develop under similar conditions \citep{Davidsson2022} and that such mantles are prone to strong thermal gradients and mechanical failure once radiative equilibrium is established at the surface \citep{alilagoaetal15,attreeetal18,Davidsson2022}.}

{An additional and complementary consequence of this delayed thermal activation is outgassing-driven spin-up of the newly exposed fragments. As sublimation intensifies several days after perihelion, asymmetric gas emission can exert substantial torques on irregularly shaped nuclei, driving rapid changes in rotation state. Rotational instability is therefore an efficient mechanism for both continued fragmentation and enhanced dust release following the initial breakup. Sub-kilometer cometary nuclei are especially vulnerable to this process, as even modest outgassing can spin them up to structural failure on timescales of days \citep{Samarasinha2013,Jewitt2021}. A well-known example is the small nucleus of 289P/Blanpain (radius $\sim$160~m), which shows only weak present-day activity yet is widely interpreted as the remnant of a rotationally driven disintegration event that produced the Phoenicid meteoroid stream \citep{Jewitt2006,Kasuga2019,Kasuga2025}.}

{For C/2025~K1, the fragment cross sections measured by HST provide only upper limits on the true nucleus sizes due to coma contamination, implying that the actual solid fragments are likely significantly smaller than the inferred kilometer-scale values. The observed near–escape-velocity separations of the fragments (0.5–10~m~s$^{-1}$), and in particular the very low relative speed of IIa and IIb ($\sim$0.5~m~s$^{-1}$), are consistent with rotational disruption rather than direct gas-drag ejection. These velocities are far below typical gas expansion speeds ($\sim$500~m~s$^{-1}$ for H$_2$O at 1~au) and are difficult to reconcile with classical gas-coupling models \citep{Whipple1951}, which predict sub–m~s$^{-1}$ ejection speeds only for meter-scale debris even at the perihelion distance of C/2025~K1. Rotational instability therefore provides a physically plausible pathway linking delayed heating, fragment spin-up, secondary breakup, and the observed dust production. In this framework, delayed heating not only regulates the onset of sublimation but also governs when rotational stresses peak, naturally coupling the timing of enhanced activity to subsequent fragmentation.}

The arclets surrounding fragment~I reinforce this interpretation. Arclet-like features have been documented in several comets \citep[e.g.,][]{Harris1997,Boehnhardt2004}, but in C/2025~K1 they are observed only in the first HST epoch (Nov.~8.56) and are absent in the subsequent images. Their fixed curvature and lack of outward motion over the paired frames within that epoch indicate short-lived, confined structures rather than propagating waves. This behavior differs markedly from the large-scale, quasi-periodic pressure waves proposed by \citet{Farnham2001} for comet C/1996~B2 (Hyakutake), where arclets appeared between the primary nucleus and a companion fragment and propagated coherently across the coma. In C/2025~K1, the arclets are anchored directly on fragment~I, do not move outward, and show no evidence for global coherence.

\citet{Harris1997} suggested that the arclets in Hyakutake were dominated by gas emission; within the STIS F28X50LP bandpass this would most plausibly arise from NH$_2$ 
($\tilde{A}^{\,2}A_{1}\;-\;\tilde{X}^{\,2}B_{1}$). Recent spectra of C/2025~K1 likewise show strong NH$_2$ emission and very weak CN, C$_2$, and C$_3$ \citep{Ganesh2025}. However, several lines of evidence argue against a gas–emission origin for the arclets in C/2025~K1. First, the features are sharply delineated, geometrically thin, morphologies inconsistent with the smoother, more diffuse structures expected from NH$_2$ fluorescence, whose parent NH$_3$ has short photochemical lifetimes at $r_h\sim 0.9$~au. Second, the arclets are tightly centered on fragment~I rather than appearing between fragments, as in Hyakutake. Third, the lack of outward expansion is incompatible with gas moving at $\sim$0.5–1~km~s$^{-1}$. Together, these properties strongly favor an interpretation in which the arclets are thin dust shells released when freshly exposed surfaces on fragment~I were first illuminated and began to warm, an origin fully consistent with the thermal-delay scenario inferred from the lightcurve.

{An important implication of this interpretation concerns the relationship between observed coma abundance ratios and the true bulk composition of the nucleus. Prior to fragmentation, strong near-surface stratification is expected, with dust-rich layers overlying deeper sublimation fronts for H$_2$O, CO$_2$, and CO. In such cases, coma abundances reflect the combined effects of temperature gradients, diffusion, and recondensation rather than the intrinsic ice abundances at depth \citep{Marboeuf2014,Davidsson2022}. Immediately after fragmentation, however, all volatile species are temporarily exposed at the surface at their bulk nucleus abundances. The observations presented here empirically constrain this window to the first $\sim$1–3~days following breakup. Any gas abundance ratios measured during this brief interval would therefore provide an unusually direct probe of nucleus interior composition. Moreover, comparisons between fragments during this early phase would offer a powerful test of whether cometary nuclei are compositionally homogeneous or heterogeneous at depth. We note that both COS and STIS spectroscopic observations that can address this were taken as part of GO-18135 and will be published separately. }

These early-time observations invite comparison with other well-studied fragmentation events. The Jupiter-family comet 73P/Schwassmann–Wachmann~3 underwent repeated episodes of fragmentation, ultimately producing dozens of pieces by 2006 \citep{Weaver2006}. Yet spectroscopic studies showed that the largest components were compositionally homogeneous \citep{DelloRusso2007}, arguing that the breakup did not expose chemically distinct reservoirs. The dynamically new comet C/1999~S4 (LINEAR) fragmented into at least 14 pieces, but by the time of its HST and VLT observations the debris field already spanned $\sim500$~square arcseconds \citep{Weaver2001}. In contrast, our HST sequence captures C/2025~K1 only days after its primary breakup, revealing just four fragments within $\lesssim30$~square arcseconds. Interestingly, both 73P and C/1999~S4 were depleted in CO, indicating that hypervolatile-driven explosions were not responsible for their breakups. C/2025~K1 similarly exhibited severe pre-perihelion depletion in carbon-bearing species, but its CO outgassing rate is not yet known.

If the observed outbursts are instead linked to rotational instabilities, simple models allow order-of-magnitude estimates of fragment sizes. Outgassing torques can spin up fragments to failure in 1--3~days for typical cometary densities ($\rho \sim 500~\mathrm{kg\,m^{-3}}$) and cohesive strengths ($S\sim10$--100~N~m$^{-2}$) \citep{Jewitt2020}. Using the observed delays between inferred breakup times and increased activity, fragments undergoing secondary disruption (such as II) are consistent with radii of roughly 5--40~m, assuming active fractions near unity. These dimensions comfortably match the slow, near–escape-velocity separation of IIa and IIb.

Finally, the activity pattern following fragmentation parallels behavior observed after cliff collapses on comet~67P/\-Chu\-ry\-u\-mov-Gerasimenko, where freshly exposed material required several rotation cycles before vigorous dust production began \citep{Pajola2017,Davidsson2023,Davidsson2024}. In both cases the thermal adjustment of interior layers governs the onset of enhanced activity. However, the fragmentation of C/2025~K1 differs fundamentally from cliff failure in scale and dynamical consequence: cliff collapses involve tens-of-meters surface failures, whereas K1 shed fragments with effective radii up to $\sim$1--2~km (upper limits given coma contamination). This indicates a nucleus-wide mechanical failure likely driven by strong perihelion heating, internal pressure buildup, or rotational torques. The cascading breakup of II into IIa and IIb aligns the event more closely with the large-fragment detachment processes described by \citet{Jewitt2022} than with small-scale surface collapse. Thus, while the subsequent activation in both contexts is shaped by the same thermal physics, the mechanical origins of the failures differ substantially.

\section{Summary}

\noindent The combined HST and LOOK observations capture C/2025~K1 (ATLAS) only days after a major breakup, providing one of the earliest and clearest views of an evolving cometary fragmentation sequence. HST resolves at least five fragments and reveals a hierarchical disruption, including a slow secondary split of fragment~II. Ground-based photometry contextualizes these events within the broader activity evolution, linking the post-perihelion 0.9~mag outburst and subsequent brightening episodes to the emergence and activation of individual fragments. Fragment motions imply that the primary breakup separated kilometer-scale pieces at a few meters per second, while the IIa--IIb split proceeded at near–escape velocity. A consistent 1–3~day delay between physical fragmentation and enhanced dust production reveals how newly exposed subsurface layers warm and begin to release volatiles. The thin, short-lived arclets around fragment~I in the first HST epoch demonstrate that the initial exposure produces only weak, surface-confined dust release, followed by stronger activity as heat diffuses into deeper, more volatile-rich layers. These patterns indicate that the volatile inventory is vertically and laterally structured: surface layers are devolatilized, while interior regions retain more abundant or more easily mobilized volatiles. The absence of immediate, explosive activity argues against hypervolatile-driven disruption and instead suggests that sublimation is throttled by the time required for vapor pressure to build within porous interior pathways. {The empirically constrained 1–3~day window between fragmentation and dust activation therefore implies a brief interval during which coma gas abundances may closely reflect bulk nucleus composition, underscoring the importance of rapid compositional measurements following breakup events.}

The morphology, kinematics, and activation behavior therefore point to a nucleus-wide mechanical failure that exposed heterogeneously mixed interior ices, followed by thermally mediated volatile release governed by diffusion and pore connectivity { and, in some fragments, by outgassing-driven rotational instability.} This behavior parallels, but greatly scales up, the delayed activation seen after cliff collapses on 67P, where the timing and morphology of dust release reflect how heat propagates, how trapped volatiles are stored, and how efficiently material is transported to the surface. These observations provide rare, time-resolved constraints on the mixing, storage, and release of volatiles in a dynamically new comet near perihelion. They show that the early evolution of a disrupted nucleus records both the mechanical architecture of the breakup and the thermal–physical response of newly exposed interior layers, offering direct insight into volatile stratification, permeability, and the efficiency of dust–gas coupling in cometary interiors.



\section*{Acknowledgements}
\noindent This work makes use of observations from the Las Cumbres Observatory global telescope network.
Observations with the LCOGT 1 m were obtained as part of the LCO Outbursting Objects Key
(LOOK) Project (Proposal IDs: LTP2025A-004).
This research is based on observations made with the NASA/ESA Hubble Space Telescope obtained from the Space Telescope Science Institute, which is operated by the Association of Universities for Research in Astronomy, Inc., under NASA contract NAS 5–26555. These observations are associated with program ID 18135 (PI D. Bodewits). We gratefully acknowledge the outstanding support of the \textit{HST}/STScI scheduling and operations staff during the execution of Program~18135. In particular, we thank Alison~Vick, Sten~Hasselquist, and TalaWanda~Monroe at the Space Telescope Science Institute for their expert assistance in obtaining these challenging observations. We would also express our gratitude to Qicheng Zhang for reporting his observations of the October outburst and fragmentation to the MPC, which helped us with the interpretation of the sequence of events.
Supporting observations were obtained by the Comet Chasers education and outreach program (\href{https://www.cometchasers.org/}{https://www.cometchasers.org/} ), led by HU, which is funded by the UK Science and Technology Facilities Council (via the DeepSpace2DeepImpact Project), the Open University and Cardiff University. It accesses the 2m, 1m and 0.4m LCOGT telescopes through the Schools Observatory/Faulkes Telescope Project (TSO2025B-01 The Schools’ Observatory/DFET), which is partly funded by the Dill Faulkes Educational Trust, and through the LCO Global Sky Partners Programme (LCOEPO2023B-013).  Observations were made by the following schools and clubs:  Institut d’Alcarras, Catalonia, Spain; Jelkovec High School, Zagreb, Croatia; Louis Cruis Astronomy Club, Brazil; St Marys Catholic Primary School, Bridgend, UK; The Coopers Company \& Coborn School, Upminster, UK (as part of the BAA Work Experience Project). Parts of this research were carried out at the Jet Propulsion Laboratory, California Institute of Technology, under a contract with the National Aeronautics and Space Administration. We acknowledge with thanks the comet observations from the COBS Comet Observation Database contributed by observers worldwide and used in this research.

During the preparation of this work the authors used ChatGPT for minor copy editing. After using this tool/service, the authors reviewed and edited the content as needed and take full responsibility for the content of the published article.
\bibliographystyle{cas-model2-names}

\bibliography{main}


\end{document}